\documentclass[traditabstract, Letter]{aa}
\usepackage     [utf8]                 {inputenc}  
\usepackage{verbatim} 
\usepackage{amsmath}
\usepackage{amssymb}
\usepackage{graphicx}
\usepackage{flafter} 
\usepackage{booktabs}
\usepackage[authoryear]{natbib}
\usepackage[colorlinks=true,linkcolor=blue,urlcolor=blue,citecolor=blue]{hyperref}
\usepackage{xspace}
\usepackage{siunitx}
\usepackage{lscape}
\usepackage{longtable}
\usepackage{threeparttable}
\usepackage{xcolor}

\makeatletter

\usepackage{txfonts}\usepackage{twoopt} 
\bibpunct{(}{)}{;}{a}{}{,} 
\usepackage{enumitem}

\usepackage[rightcaption]{sidecap}
\usepackage{graphicx} 
\graphicspath{ {images/} }

\titlerunning{Reconstruction of annual solar irradiance over the last three millennia}
\authorrunning{D.~Temaj}
\makeatother
\begin{document}

\title{Reconstruction of annual solar irradiance over the last three millennia}
\author{D.~Temaj\inst{\ref{MPS}, \ref{Braunschweig}}\thanks{temaj@mps.mpg.de},
N.A.~Krivova\inst{\ref{MPS}}, 
S.K.~Solanki\inst{\ref{MPS},\ref{Korea}},
I.G. Usoskin\inst{\ref{Finland}},
T. Chatzistergos\inst{\ref{MPS}}
}
\institute{
Max-Planck-Institut für Sonnensystemforschung, Justus-von-Liebig-Weg 3, Göttingen, Germany\label{MPS}
\and Technical University of Braunschweig, Hans-Sommer-Straße,
Braunschweig, 3329, Germany \label{Braunschweig}
\and School of Space Research, Kyung Hee University, Yongin, Gyeonggi 17104, Republic of Korea \label{Korea}
\and Space Physics and Astronomy Research Unit and Sodankyl\"a Geophysical Observatory, University of Oulu, Finland \label{Finland}
 }
\date{\today}
\abstract{Solar irradiance measurements are limited to the last few decades, requiring reconstructions to assess solar variability on longer timescales and its impact on Earth's climate. We present the first physics-based reconstruction of total solar irradiance (TSI) at annual resolution over the last three millennia. The reconstruction is obtained by extending the SATIRE-T model beyond the telescopic era using recently published, annually resolved sunspot number series derived from cosmogenic isotope records. This yields a continuous, physics-based TSI record extending from the satellite era back over the last three millennia, with annual resolution throughout the pre-telescopic period. Over the full three-millennia interval, the reconstructed TSI exhibits a maximum difference of $1.04_{-0.2}^{+0.14}\,\mathrm{W\,m^{-2}}$, defined as the difference between the maximum and minimum of the 50-yr running mean values.}
\maketitle
%
%

\section{Introduction}

The Sun is the dominant source of energy for the Earth system \citep[e.g.,][]{kren_where_2017}, and variations in its radiative output influence the climate on various timescales \citep[e.g.,][]{haigh_sun_2007,gray_solar_2010,solanki_solar_2013}.
The solar radiative flux received at a distance of 1~AU is termed solar irradiance.
Its spectrally integrated value is known as the total solar irradiance (TSI), while its wavelength-dependent distribution is referred to as the spectral solar irradiance (SSI).
Continuous space-based measurements of TSI have been available only since 1978, revealing variability on timescales from minutes to decades \citep[e.g.,][]{kopp_solar_2025}. To understand solar variability on longer timescales and assess its impact on Earth's climate, irradiance reconstructions extending beyond the satellite era are therefore required.

Variations in solar irradiance are driven by the evolution of the solar surface magnetic field through the competing effects of dark sunspots on the one hand and bright faculae and network regions on the other \citep[e.g.,][]{krivova_reconstruction_2003,yeo_solar_2017,shapiro_nature_2017}.
Consequently, irradiance reconstructions require knowledge of the temporal evolution of solar surface magnetism.
Direct observations of the solar magnetic field are available only for the past few decades \citep[e.g.,][]{pevtsov_long-term_2021}, and proxies of solar magnetic activity must therefore be employed to extend irradiance records further back in time.

Sunspot numbers (SN) provide the longest direct observational record of solar activity, extending over about four centuries \citep[see reviews by][]{arlt_historical_2020,clette_recalibration_2023}.
However, sunspots represent only the dark component of solar surface magnetism and provide no direct information on the bright magnetic features~-- faculae and network~-- that dominate irradiance variability on the solar-cycle and longer time scales.
Consequently, irradiance reconstructions based solely on sunspot data must infer the evolution of the total surface magnetic field indirectly, under additional assumptions \citep[see, e.g.,][for reviews]{solanki_solar_2013,chatzistergos_long-term_2023,chatzistergos_understanding_2024}.
This limitation is present at all activity levels, but becomes particularly critical during prolonged periods of very low activity, when the SN is close to zero over extended periods of time.
During such intervals, sunspot-based models lose information on magnetic flux emergence, leading to increased uncertainty in the reconstructed irradiance variability.

To address this limitation, \cite{krivova_modelling_2021} revised the magnetic flux evolution model originally developed by \cite{solanki_evolution_2000,solanki_search_2002} by introducing a more realistic description of the emergence of small-scale magnetic regions, constrained by modern observations \citep{thornton_small-scale_2011}. 
Employed within the SATIRE (Spectral And Total Irradiance REconstructions; \citealp{krivova_towards_2011,solanki_solar_2013}) model, this revised version enables physically consistent reconstructions of solar magnetic flux and irradiance over the telescopic era when driven by sunspot number records.
A recent implementation and optimisation of this approach, including extensive validation against independent observations and reconstructions of magnetic flux and irradiance, is presented by \citet[][hereafter Paper 1]{temaj_solar_2026}.

On longer timescales, both physics-based \citep{vieira_evolution_2011,wu_solar_2018} and empirical, often regression-based
\citep{steinhilber_9400_2012,roth_reconstruction_2013,penza_reconstruction_2024} irradiance reconstructions extending over multiple millennia, up to the entire Holocene period, have been obtained using cosmogenic isotope records.
Owing to the low signal-to-noise ratio inherent in the underlying radionuclide data, such reconstructions were limited to decadal resolution. 
Recent advances in the measurements and analysis of cosmogenic isotope data \citep{brehm_eleven-year_2021,brehm_tracing_2025, Heaton24,wang_patterns_2026} have led to the publication of annually resolved SN reconstructions covering the last three millennia \citep{usoskin_solar_2021, usoskin_sunspot_2025, usoskin_sunspot_2026}.
These datasets provide, for the first time, continuous solar activity estimates at annual cadence well beyond the telescopic era.

Here, we use these reconstructions to extend the SATIRE-T model beyond the period of direct solar observations.
This yields the first physics-based, annually resolved reconstruction of solar irradiance for the pre-telescopic era, spanning the last three millennia and extending the satellite- and telescopic-era irradiance record by more than 2.5 millennia.

\section{Model}
\label{sec:model}
\subsection{Modelling approach}
\label{sec:approach}

We reconstruct solar irradiance using the SATIRE-T model as implemented, revised, and optimised in Paper 1. The model yields both TSI and SSI. In the present Letter, we focus on the long-term variability of TSI. The SSI time series are produced consistently using the same methodology as in the telescopic-era reconstruction and are provided to the community for use, but they are not discussed separately here. 

In SATIRE-T, solar irradiance variability is computed from the fractional disc coverage by sunspots, faculae, and network and their radiative contrasts relative to the quiet Sun. 
The surface coverages are derived from the temporal evolution of the solar surface magnetic field inferred from SN input. This evolution is described by a set of coupled ordinary differential equations that account for magnetic flux emergence, transport, and decay.

The implementation employed here follows the revised magnetic-flux evolution model of \citet{krivova_modelling_2021}, in which magnetic regions of all sizes emerge according to a single power-law distribution in magnetic flux, consistent with observations of small-scale magnetic features \citep{thornton_small-scale_2011}.
The slope of the distribution depends on the level of solar activity, quantified by the sunspot number.
This formulation ensures a continuous emergence of small-scale magnetic flux even during periods of very low or zero SN, thereby maintaining a realistic background magnetic field during extended minima.

The reconstructed magnetic fluxes determine the fraction of the solar disc covered by different magnetic components, here called filling factors. These filling factors are converted into irradiance variations using pre-computed intensity spectra for each component \citep{unruh_spectral_1999}, following the standard SATIRE formalism.

All model parameters are adopted directly from Paper 1, without re-optimisation or re-tuning. The present work, therefore, differs from the telescopic-era reconstruction only in the SN input used to drive the model.

\subsection{Sunspot number input and temporal resolution}
\label{sec:sn}

SATIRE-T requires SN input at daily cadence. For the telescopic era, we use the same SN datasets as in Paper~1, which are employed here solely for validation and comparison with the pre-telescopic reconstructions.

The primary telescopic SN input is the International Sunspot Number version 2 \citep[ISNv2;][]{clette_recalibration_2023}, providing daily, monthly, and annual values back to 1818, 1749, and 1700, respectively. Daily ISNv2 values are used directly where available, while monthly and annual data prior to 1818 were interpolated to daily cadence. To illustrate the spread among historical sunspot observations, we also considered the Group Sunspot Number (GSN) series by \citet[][CEA17; which we scaled to the level of ISNv2]{chatzistergos_new_2017}, which begins in 1739 but is sparse prior to 1749 and is therefore not used before 1749.

To extend telescopic SN records further back in time, we additionally employed the GSN series by \citet[][HoSc98]{hoyt_group_1998}, which provides daily values from 1610 onward.
Due to misclassifications of zero-group days during the Maunder Minimum and other criticisms of this series \citep[e.g.][]{vaquero_revised_2016,clette_recalibration_2023,chatzistergos_assessment_2025}, HoSc98 is used only prior to 1749 and is treated as a lower-limit estimate of solar activity.
The HoSc98 series is scaled separately to ISNv2 and CEA17 and used solely where no more reliable telescopic data are available.

As an alternative constraint on solar activity during the Maunder Minimum, we also consider annual SN estimates derived using the active-day fraction method by \citet[][]{carrasco_relationship_2022,carrasco_understanding_2024} over 1635--1721.
These data provide an approximate upper-limit estimate of solar activity during the Maunder Minimum (although before 1650 and over roughly 1710--1720 they are slightly higher than our ISNv2 compilation).
Over the period 1722--1748, not covered by the \citet[][]{carrasco_relationship_2022,carrasco_understanding_2024} data, we use annual ISNv2 interpolated to daily values.

In addition to the telescopic SN record, we employ published, annually resolved SN reconstructions derived from cosmogenic isotope data. Specifically, we use the reconstructions by \citet{usoskin_solar_2021,usoskin_sunspot_2025,usoskin_sunspot_2026}, covering the periods 971--1899\,CE, 997--1\,BCE, and 1--969\,CE, respectively. These reconstructions are based on improved analyses of $^{14}$C isotope records from tree rings \citep{brehm_eleven-year_2021,brehm_tracing_2025,wang_patterns_2026}. Together, these datasets provide continuous estimates of solar activity at annual resolution over the last three millennia, enabling a consistent extension of the SATIRE-T model beyond the telescopic era.
A one-year (970 CE) and a six-year (6 BCE - 1 BCE) gaps in the datasets were filled by interpolation.

Because the magnetic flux evolution model operates at a daily cadence, the annually resolved SN must be converted to a daily time series.
For this purpose, we adopt a box-car representation, in which the annual SN is assigned uniformly to all days of the corresponding year, thereby preserving the annual mean exactly.
Tests performed using the telescopic-era sunspot record demonstrate that this procedure preserves the annually averaged irradiance variability and does not introduce systematic biases relative to reconstructions driven by the original daily sunspot data (see Appendix~\ref{app:Method-validation}).
During periods of extremely low activity, such as grand minima, this procedure may slightly overestimate the persistence of emerged flux on the surface, as any short-lived emergence within a given year is represented as a weak but continuous contribution throughout the year.

\begin{SCfigure*}
    \centering
    \includegraphics[width = 0.7\textwidth]{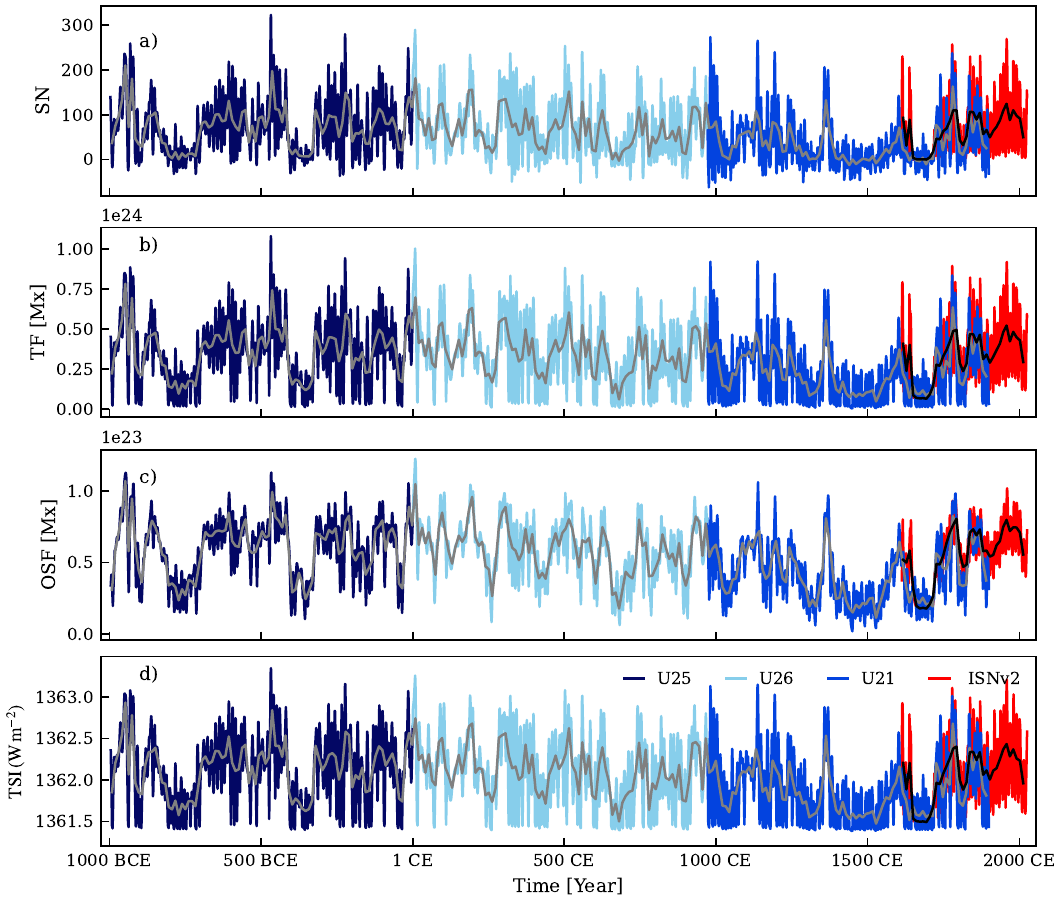}
    \caption{ (a) Annual sunspot numbers (ISNv2 in red and $^{14}$C-based SN in different shades of blue, as indicated in the legend -- U21, U25, and U26 stand for \citealt{usoskin_solar_2021,usoskin_sunspot_2025,usoskin_sunspot_2026}). 
     The different blue shades correspond only to the origin of the input data and do not reflect separate reconstructions.
    These SN values are used as input to reconstruct:
   (b) Total magnetic flux,
   (c) Open solar magnetic flux, and
   (d) Total solar irradiance.
11-year means for the $^{14}$C- and ISNv2-based reconstructions are shown in grey and black, respectively.
}
\label{fig: Reconstructions_full_period}
\end{SCfigure*}

\begin{figure}
        \includegraphics[width = 0.5\textwidth]{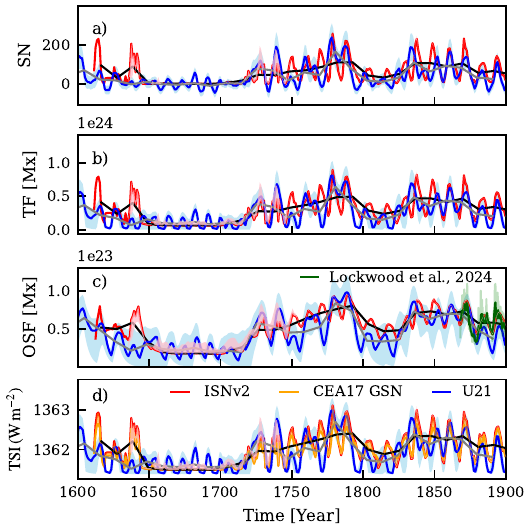}
    \caption{
    Same as Fig.~\ref{fig: Reconstructions_full_period} for the 1600--1900 overlap period of $^{14}$C-derived and telescopic SN. 
    Panel (c) also shows \cite{lockwood_reconstruction_2024} OSF at Carrington-rotation (light green) and three–Carrington-rotation (dark green) resolutions.
    Panel (d) also includes TSI from Paper~1 based on CEA17 GSN (orange). 
    The range between the ISNv2- and CEA17-based reconstructions is shaded light orange, that using ISNv2 extension with HoSc98 and \citet[][see Sect.~\ref{sec:sn}]{carrasco_understanding_2024} is shaded light pink. 
    Blue shading indicates the $^{14}$C-based SN input uncertainty, obtained by driving the model with the upper and lower bounds of the SN series from \citet{usoskin_solar_2021}.
    }
\label{fig: Reconstructions_zoom}
\end{figure}

\subsection{Magnetic flux and irradiance computation}

From the daily SN input, the magnetic flux evolution model computes the emergence, transfer, and decay of magnetic flux, yielding time series of total surface magnetic flux and open magnetic flux. 
Sunspot areas are obtained from the SN using the same empirical relationship as adopted in the telescopic-era reconstruction of Paper 1. 
The filling factors of faculae and network regions are derived from the reconstructed magnetic fluxes.

Total solar irradiance is then computed following the SATIRE prescription, using identical assumptions, parameters, and contrasts for both the telescopic and pre-telescopic periods.
As a result, the present reconstruction constitutes a direct and physically consistent extension of the validated telescopic-era SATIRE-T reconstruction into the pre-telescopic period.

\section{Results and discussion}
\label{sec:res}

Figure~\ref{fig: Reconstructions_full_period}b)-d) presents the
total (TF) and open (OSF) magnetic fluxes, and TSI over the last three millennia based on annually resolved $^{14}$C-derived SNs \citep{usoskin_solar_2021,usoskin_sunspot_2025,usoskin_sunspot_2026}, which are shown in panel a.
The reconstruction exhibits pronounced variability on solar-cycle to centennial timescales, including extended intervals of significantly reduced activity commonly referred to as grand minima.
To quantify the secular TSI variability, we considered 50-yr running means, consistent with the definition adopted in Paper 1.
Over the full three-millennia interval, the reconstructed TSI exhibits a maximum difference of about $1.04\,\mathrm{W/m^2}$ (with the uncertainty range being 0.84--1.18$\,\mathrm{W/m^2}$) between the maximum and minimum values of the 50-yr running means.
The quoted range reflects the spread obtained by driving the model with the upper and lower bounds of the $^{14}$C-based SN reconstruction, provided in the source papers.
This range is somewhat larger than the 0.67--0.75$\,\mathrm{W/m^2}$ increase between the Maunder Minimum and the present day obtained for the telescopic era (Paper 1).

The amplitude of the secular variability inferred here lies at the lower end of previously published multi-millennial reconstructions.
Appendix~\ref{app:comparison} provides a quantitative comparison with selected earlier reconstructions.
However such comparisons are inherently limited by differences in temporal resolution, while the magnitude of long-term variability is effectively constrained by the model formulation and its calibration to modern observations (e.g., TSI and/or neutron monitor data), rather than being independently constrained by the millennial extension.

Figure~\ref{fig: Reconstructions_zoom} shows the period of overlap between the $^{14}$C-based and direct telescopic SN records in greater detail.
The spread between reconstructions driven by alternative telescopic sunspot datasets illustrates the uncertainty in the inferred magnetic fluxes and irradiance variability arising from ambiguities in historical sunspot observations, particularly during the early telescopic era.
The $^{14}$C-based TSI reconstruction lies well within this range, indicating that extending SATIRE-T beyond the telescopic era using $^{14}$C-derived SN does not introduce inconsistencies in the magnetic activity--irradiance relationship.
This is further confirmed by Fig.~\ref{fig:tsi_vs_sn} in Appendix~\ref{app:TSI-comparison}, showing the relationship between the SN and the reconstructed TSI for both telescopic and pre-telescopic periods.
The reconstructed OSF is also consistent with independent reconstructions based on geomagnetic indices \citep{lockwood_reconstruction_2024} during the overlap period (Fig.~\ref{fig: Reconstructions_zoom}c).

The relationship between the SN and the reconstructed TSI for the pre-telescopic period closely matches that obtained for the telescopic era for non-negative SN (Appendix~\ref{app:TSI-comparison}). However, a notable feature of the $^{14}$C-based SN reconstructions is that annual values occasionally become slightly negative during periods of very low activity. Such negative values are not physical in a literal sense but arise from the statistical nature of the reconstruction and the noise inherent in isotope records. Within the uncertainties of the reconstructions \citep{usoskin_solar_2021,usoskin_sunspot_2025,usoskin_sunspot_2026}, they are fully consistent with zero SN. 
The mean level of these fluctuations in the $^{14}$C-based SN nevertheless reflects the overall modulation of cosmic rays and thus the underlying level of solar magnetic activity, even if it falls below the sunspot formation threshold.
We therefore retained the negative values in the model input.
Because for SN$<$0, the activity-dependent slope of the flux-emergence distribution lies outside the observationally calibrated range (see Sect.~\ref{sec:approach} and Paper~1), we linearly extrapolated it from the lowest positive values. As this regime corresponds to vanishingly weak flux emergence, this choice has a negligible impact on the reconstructed irradiance.

Through the non-linear response of the model at very low activity levels, these
fluctuations propagate into the reconstructed magnetic fluxes and irradiance.
The resulting short-term variability during extended minima should therefore not be interpreted as resolved solar cycles. 
Rather, the reconstruction primarily constrains the mean activity level during such periods.

In addition, because the irradiance approaches a background level corresponding to the absence of large-scale flux emergence (reached for slightly negative SN values, see Fig.~\ref{fig:tsi_vs_sn} in Appendix~\ref{app:TSI-comparison}), fluctuations symmetric about SN=0 do not translate into symmetric variations in TSI. 
Positive excursions around zero increase TSI more effectively than negative excursions reduce it.
This asymmetry can lead to a slight upward bias in the time-averaged TSI during grand minima, even when negative SN values are retained in the model input.

\section{Conclusion}
\label{sec:conclusion}

We have presented the first annually resolved reconstruction of total solar irradiance extending more than 2.5 millennia prior to the telescopic era, obtained by applying the revised SATIRE-T model to published annual sunspot number reconstructions derived from cosmogenic isotopes.
While physics-based irradiance reconstructions over the Holocene exist at decadal resolution \citep{vieira_evolution_2011,wu_solar_2018}, the present work demonstrates that annually resolved irradiance variability can be reconstructed in a physically consistent manner beyond the period of direct solar observations.
Over the full three-millennia interval, the reconstructed TSI exhibits a maximum difference of 
$1.04_{-0.2}^{+0.14}\,\mathrm{W\,m^{-2}}$, defined as the difference between the maximum and minimum of the 50-yr running means.
This reconstruction provides an important record for studies requiring long-term, annually resolved solar forcing.

\section{Data availability}
TSI and SSI reconstructions are available at \url{https://www2.mps.mpg.de/projects/sun-climate/data.html}, and through the CDS via \url{http://cdsarc.u-strasbg.fr/}.
\begin{acknowledgements}
D.T. was supported through the International Max-Planck Research School (IMPRS) for Solar System Science at the Technical University of Braunschweig.
S.K.S and T.C. acknowledge ERC funding under the EU Horizon 2020 program (grant No.~101097844 — project WINSUN).
This research has made use of the Astrophysics Data System Bibliographic Services, funded by NASA under Cooperative Agreement 80NSSC21M00561.
\end{acknowledgements}

\bibliographystyle{aa}
\bibliography{references}

\newpage

\begin{appendix}
\section{Validation of the annual-to-daily SN interpolation}
\label{app:Method-validation}

\begin{figure}[!htb]
 \includegraphics{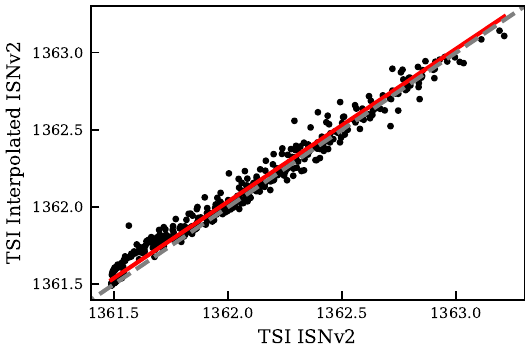}
  \caption{Comparison between the TSI reconstruction from Paper~1 based on the daily ISNv2 dataset and the reconstruction obtained by interpolating annual mean ISNv2 values to daily cadence, both shown as annual means. 
  The ordinary least-squares bisector fit is shown in red, while the gray dashed line shows the one-to-one relation. } 
  \label{fig:scatter_TSI_vs_TSI}
\end{figure}

As a validation test of the annual-to-daily resampling procedure, Figure~\ref{fig:scatter_TSI_vs_TSI} compares the TSI reconstruction obtained by driving the SATIRE-T model with the original daily ISNv2 sunspot number series (Paper~1) to a reconstruction in which the daily sunspot numbers were first averaged to annual means and then converted back to daily cadence using the box-car representation adopted in the present work.
We find a Pearson correlation coefficient of $R_c$ = 0.99 between the two annually averaged TSI reconstructions, and an ordinary least-squares bisector fit \citep{isobe_linear_1990} between them gives a slope of 0.99. 

The close agreement and near-unity slope demonstrate that the resampling procedure preserves the annually averaged irradiance variability and does not introduce systematic offsets.
This confirms that SATIRE-T reconstructions driven by annually resolved sunspot numbers converted to daily cadence using a box-car representation are directly comparable to reconstructions based on true daily sunspot data.

\section{Comparison to other multi-millennial reconstructions}
\label{app:comparison}

For comparison with earlier multi-millennial irradiance reconstructions \citep[including][]{wu_solar_2018, egorova_revised_2018, lean_estimating_2018,roth_reconstruction_2013, muscheler_solar_2007, delaygue_antarctic_2011, penza_reconstruction_2024, steinhilber_9400_2012}, we first homogenise the temporal resolution of all series, to the extent possible. The reconstruction by \citet{steinhilber_9400_2012} is provided at a cadence of 22~years, while \citet{wu_solar_2018} has decadal resolution. All other reconstructions are available at annual sampling, although solar-cycle variability is generally not independently reconstructed. To enable a consistent comparison, we therefore represent all series as 22-year running means (and 2-point running means for decadal data).

For a quantitative comparison and for consistency with Paper~1, we define the secular variability as the difference between the maximum and minimum of the 50-year running means. This is computed over the full overlapping period with our reconstruction (996\,BCE--1900\,CE), where available, and additionally over the last millennium (850--1850\,CE), when all datasets are available.

\begin{figure*}
    \includegraphics{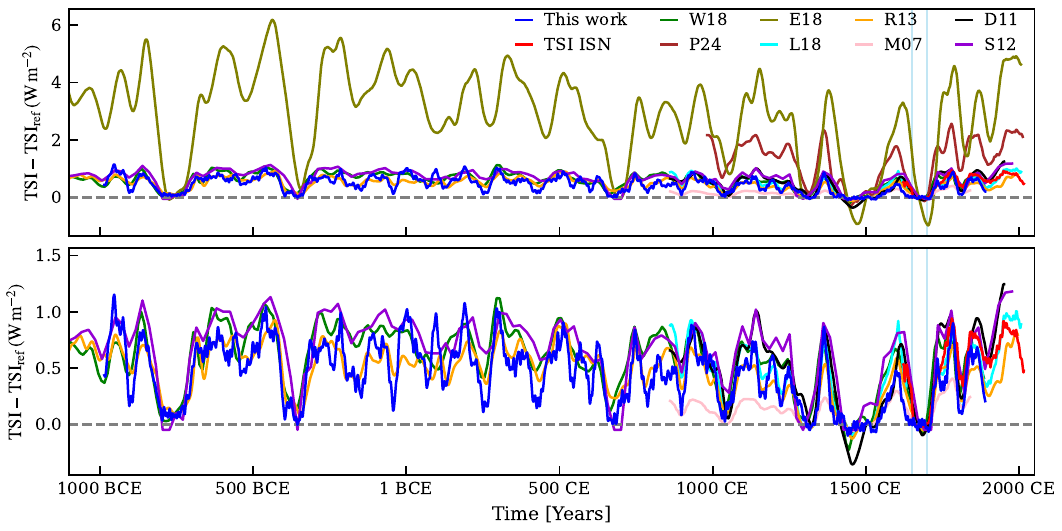}
    \caption{Comparison of multi-millennial TSI reconstructions shown relative to their mean level over the Maunder Minimum reference period (1650--1700) marked by the vertical blue lines.
   Shown are our reconstruction (blue), the telescopic sunspot-based reconstruction of \citet[][red; TSI ISN]{temaj_solar_2026}, and reconstructions by 
\citet[][green; W18]{wu_solar_2018};
\citet[][brown; P24]{penza_reconstruction_2024};
\citet[][olive; E18]{egorova_revised_2018};
\citet[][light blue; L18]{lean_estimating_2018};
\citet[][orange; R13]{roth_reconstruction_2013};
\citet[][pink; M07]{muscheler_solar_2007};
\citet[][black; D11]{delaygue_antarctic_2011};
and \citet[][purple; S12]{steinhilber_9400_2012}.
All series are shown as 22-year moving means, except for \citet[][]{wu_solar_2018}, which has decadal resolution and is therefore represented by a 20-year moving mean, while
\citet[][]{steinhilber_9400_2012} is provided at 22-year cadence.
The lower panel excludes the reconstructions by \citet{penza_reconstruction_2024} and \citet{egorova_revised_2018} to improve the visibility of the remaining series.
The horizontal dashed grey line marks $\Delta$TSI$=0$. 
}\label{fig:tsi_comparisons}
\end{figure*}

For visual comparison (Fig.~\ref{fig:tsi_comparisons}), all series are shown relative to their mean TSI value over the reference period 1650--1700, roughly corresponding to the Maunder Minimum. We choose this period as reference as it is covered by all reconstructions, while simultaneously allowing an immediate rough visualisation of the amplitudes of the secular variability in different models. 

Overall, the amplitude of secular variability in our reconstruction lies at the lower end of the published range (Fig.~\ref{fig:tsi_comparisons}, Table~\ref{tab:TSI_variability}). Differences in long-term variability between reconstructions in the pre-telescopic era arise from both the underlying model architecture and the way irradiance variability is linked to modern observations. In the present work, the model parameters were calibrated in Paper~1 using the telescopic sunspot-number-based reconstruction, which directly overlaps with satellite TSI measurements. As a result, the amplitude of secular variability is effectively set by the model formulation and its calibration over the telescopic period, which remain unchanged when the model is driven by the $^{14}$C-based sunspot reconstruction.

While approaches employed by other models differ in formulation, they are likewise constrained by the modern observational period.
Independently of their architecture, they either regress cosmogenic isotope-derived quantities to TSI directly or calibrate the modulation potential to neutron monitor data before converting it to irradiance. In both cases, the long-term irradiance changes are likewise anchored to the more recent period. Consequently, the spread in secular variability across reconstructions largely reflects differences already present in the telescopic era, whereas the millennial extension primarily broadens the covered activity range over which these differences manifest themselves.

\begin{table}[ht]
\centering
\caption{Comparison of the amplitude of secular TSI variability for different reconstructions.}
\begin{tabular}{lcc}
  \hline
  \hline
  Dataset & \multicolumn{2}{c}{$\Delta \mathrm{TSI}\,[\mathrm{W/m^2}]$}\\
  & 3 millennia & 1 millennium \\
  \hline
  This Work & $1.04_{-0.2}^{+0.14}$           & $0.8_{-0.23}^{+0.11}$ \\
  
  \cite{wu_solar_2018}          & $1.30$   & $0.94$ \\
  \cite{egorova_revised_2018}     & $6.62$   & $4.80$ \\
  \cite{lean_estimating_2018}     & --       & $0.94$\\
  \cite{roth_reconstruction_2013} & $0.95$   & $0.72$\\
  \cite{muscheler_solar_2007}     & --       & $0.41$\\
  \cite{delaygue_antarctic_2011}  & --       & $1.23$\\
  \cite{penza_reconstruction_2024}& --       & $2.27$\\
  \cite{steinhilber_9400_2012}    & $1.23$   & $0.94$\\
  \hline
\end{tabular}\label{tab:TSI_variability}
\tablefoot{The amplitude is defined as the difference between the maximum and minimum values of the 50-yr running mean. \cite{wu_solar_2018} has a decadal resolution, so we apply a 5-decade running means, while \cite{steinhilber_9400_2012} has a 22-yr resolution, for which we apply 2-point running means. Values are given for the full period overlapping with our reconstruction (996\,BCE--1900\,CE), where available, and for the last millennium (850--1850\,CE), during which all reconstructions are available.} 
\end{table}

\medskip
\section{Consistency of the TSI reconstructions from different sunspot number inputs}
\label{app:TSI-comparison}

\begin{figure}
   \includegraphics{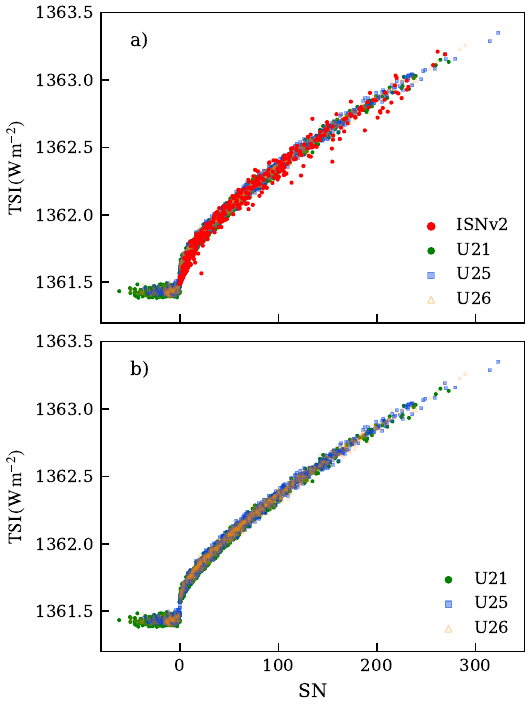}
    \caption{Relationship between reconstructed annual TSI and the corresponding sunspot number input.
    Top panel: Reconstruction based on telescopic SN (ISNv2; red) and on $^{14}$C-derived SN series \citep[][
    green, blue, and orange, respectively, indicating the different input datasets]{usoskin_solar_2021, usoskin_sunspot_2025, usoskin_sunspot_2026}.
    Bottom panel: Same as top panel, but showing only the reconstruction based on the $^{14}$C-derived SN series.}
    \label{fig:tsi_vs_sn}
\end{figure}

Figure~\ref{fig:tsi_vs_sn} illustrates the relationship between reconstructed TSI and the corresponding SN input. 
The top panel compares the reconstruction based on telescopic SN (ISNv2) with those based on
$^{14}$C-derived SN series, where the different isotope-based datasets are indicated by different colours (green, blue, and orange). 
The bottom panel shows the same relationship, but comparing only the $^{14}$C-based reconstructions.

For non-negative sunspot numbers, all reconstructions follow the same relationship in both shape and amplitude. This demonstrates the consistency of the TSI reconstructions from different SN inputs and shows that extending SATIRE-T from telescopic to cosmogenic-isotope-based sunspot numbers preserves the same dependence.
The different $^{14}$C-based datasets overlap closely over their common range of activity and exhibit no systematic differences in the SN--TSI relationship. The mean TSI levels inferred from the different $^{14}$C-based input series agree within $\lesssim $0.3\,W\,m$^{-2}$, with comparable variances. Small differences are consistent with variations in the underlying activity levels, because each dataset samples distinct intervals of solar activity. Importantly, no systematic offsets between the datasets are evident.

The $^{14}$C-based reconstruction additionally includes periods with slightly negative annual sunspot numbers, which occur during episodes of extremely low solar activity. In the irradiance reconstruction, these points form a narrow branch clustered near the low-activity irradiance plateau, as expected under grand-minimum conditions. This is because the emergence rate during such periods is vanishingly low.

\end{appendix}

\end{document}